\documentclass[aps,prl,twocolumn,superscriptaddress,showpacs]{revtex4}
\usepackage{amsmath}
\usepackage{graphicx}

\begin{document}

\title{Phase analysis of quantum oscillations in graphite}
\author{Igor A. Luk'yanchuk}

\affiliation {University of Picardie Jules Verne,
  Laboratory of Condensed Matter Physics, Amiens, 80039, France}
\affiliation {L. D. Landau Institute for Theoretical Physics,
Moscow, Russia}

\author{Yakov Kopelevich}

\affiliation{ Instituto de Física "Gleb Wataghin", Universidade
Estadual de Campinas, Unicamp 13083-970, Campinas, Sao Paulo,
Brazil}

\date{\today}

\begin{abstract}
The quantum de Haas van Alphen (dHvA) and Shubnikov de Haas (SdH)
oscillations measured in graphite were decomposed by pass-band
filtering onto contributions from three different groups of
carriers. Generalizing the theory of dHvA oscillations for 2D
carriers with arbitrary spectrum and by detecting the oscillation
frequencies using a method of two-dimensional phase-frequency
analysis which we developed, we identified these carriers as (i)
minority holes having a 2D parabolic massive spectrum
$p_{\perp}^2/2m_\perp$, (ii) massive majority electrons with a 3D
spectrum and (iii) majority holes with a 2D Dirac-like spectrum
$\pm vp_{\perp }$ which seems to be responsible for the unusual
strongly-correlated electronic phenomena in graphite.

\end{abstract}


\maketitle

Studies of electronic properties of graphite have considerably
intensified during the past decade because of the discovery of novel
carbon-based materials such as fullerenes and nanotubes constructed
from wrapped graphite sheets \cite{Saito1998}. The quasi 2D
conductivity of graphite occurs mostly inside the carbon layers due
to hexagonal networks of overlapped $\pi -$bonds. In this
single-layer approximation the Fermi surface (FS) is reduced to two
points at the opposite corners of the 2D hexagonal Brillouin zone
where the valence and conducting bands touch each other leading to
the Dirac cone spectrum $E(p) =\pm vp_{\perp }$, and the charge
carriers are described by the massless (2+1) dimensional Dirac
fermions \cite{Gonzalez96,Abrikosov99,Khveshchenko01}. This
point-like spectrum singularity and strong Coulomb coupling between
fermions are assumed to be responsible for unusual electronic
features in graphite such as, e. g., experimentally observed
magnetic-field-driven metal-insulator transition
\cite{Kopelevich2003PRL,Kopelevich2003ASS}.

However, to the best of our knowledge no unambiguous experimental
evidence for Dirac fermions  in graphite has been yet reported. In
real graphite samples the inter-layer hopping leads to
$p_z$-spectrum dispersion with opening of cigar-like FS pockets
elongated along the corner edge H-K-H of the 3D hexaedronal
Brillouin zone. The discussed in detail in Refs.
~\cite{Brandt1988,Soule1964,Williamson1965} 3D FS has a
complicated multi-sheet structure and provides for the different
groups of carriers. Band calculations show that in addition to the
two principle majority groups of electrons (e) and holes (h) which
are located close to points K and H of the Brillouin zone, several
minority (m) low-concentration groups carriers are possible. The
nature and location of the minority pockets are very sensitive to
the parameters of the band structure calculations and to the
crystalline disorder.

One can expect that the model of strongly interacting 2D Dirac
fermions is applicable to the real quasi 2D FS in graphite since
the Dirac singularity is the topological property of the
electronic spectra \cite{Mikitik99} that should be stable towards
the weak 3D inter-layer coupling.

In this Letter we make a comparative phase analysis of quantum de
Haas van Alphen (dHvA) oscillations of the magnetization $M(H)$
and of Shubnikov de Haas (SdH) oscillations of the resistance
$R(H)$ which provides  direct evidence that the group of carriers,
associated with majority holes (h) has the Dirac singularity in
the spectrum. Other groups: (e and m) have the massive spectrum
$E(p)=p_{\perp }^2/2m_\perp$.

Quantum dHvA and SdH oscillations are the appropriate tools to
study FS properties and to
distinguish between different types of fermion carriers.
Early measurements of dHvA and SdH
oscillations \cite
{Brandt1988,Soule1964,Williamson1965,Woollam1971} in
agreement with band structure calculations demonstrated that
two majority (e and h) and at least one minority (m) group of
carriers exist in graphite.

To discriminate between normal, i. e. described by the massive
spectrum, and Dirac fermions, we explored the rarely measured
\textit{phase} of quantum oscillations. Generally, phase detection
encounters  difficulties related to interference of the
contributions from different groups of carrier and to its
sensitivity to errors in frequency determination. We overcame the
problem using a specially-developed two-dimensional phase-frequency
analysis of its Fourier image. We distinguish also the different
carrier groups by applying selective pass-band filtering of the
oscillating signal.

\begin{figure}[!bt]
\centering
\includegraphics [width=5.5cm] {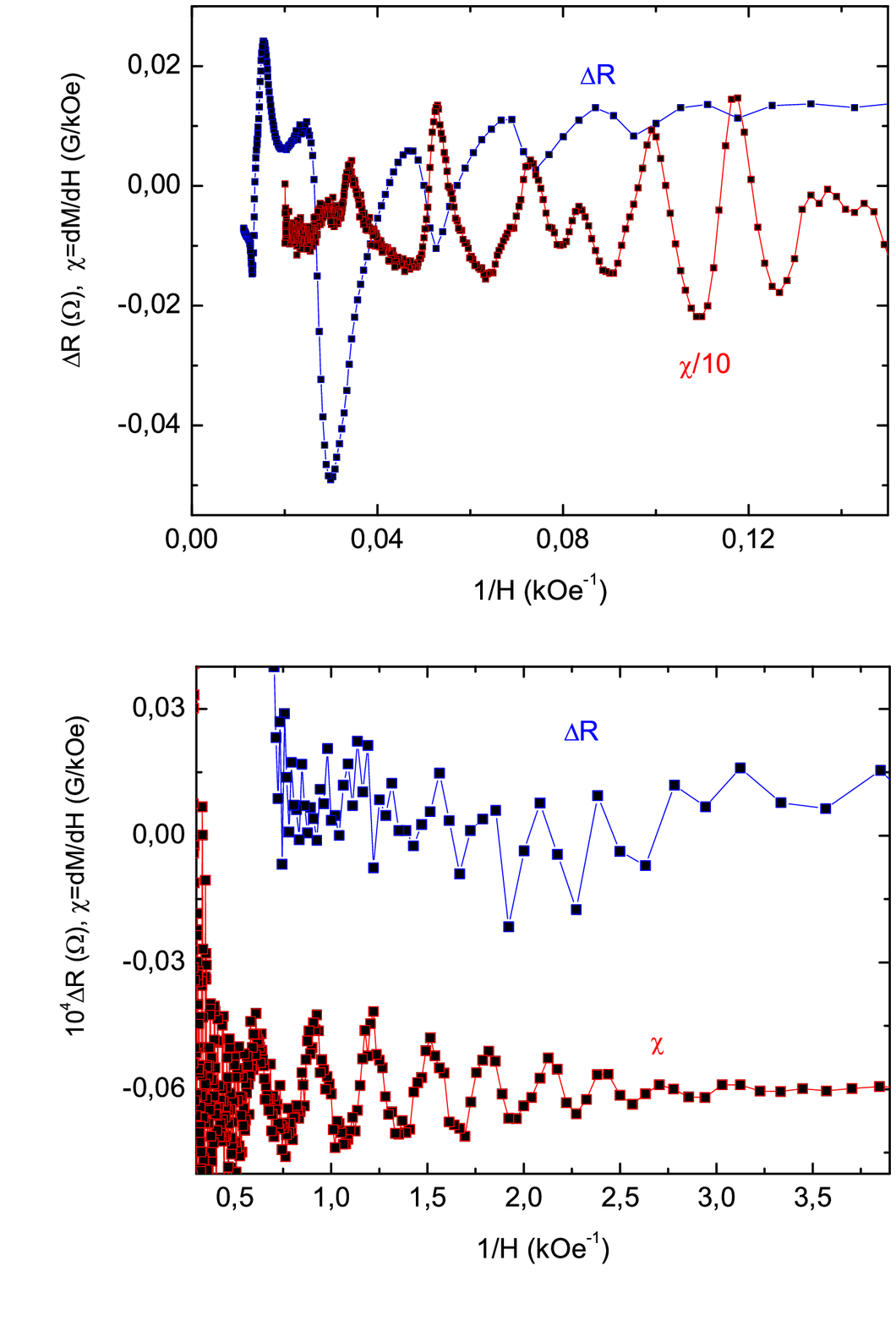}
\caption{dHvA and SdH oscillations in graphite. Upper panel shows the region
of fields $7kOe<H<50kOe$, characteristic for majority oscillations whereas
the region of fields in low panel $0.25kOe<H<2.5kOe$ corresponds to minority
oscillations}
\label{DiracFigExperiment}
\end{figure}

The magnetoresistance $R(H)$\ and magnetization $M(H)$\ data were
obtained on well-characterized highly oriented pyrolytic graphite
(HOPG) sample from the Union Carbide Co (HOPG-UC) as described in
Ref. \cite {Kopelevich2003PRL}. Briefly, low-frequency ($1$\ Hz)
and dc standard
four-probe magnetoresistance measurements were performed in magnetic field $%
0-90$\ kOe applied parallel to the sample hexagonal c-axis ($H\parallel c$),
and at the lowest available temperature $T=2$\ K using Quantum Design
PPMS-9T and Janis-9T magnet He-cryostats. Magnetization measurements $M(H)$\
were carried out with $H\parallel c$\ by means of the SQUID magnetometer
MPMS5 (Quantum Design).

\begin{figure}[!bt]
\centering
\includegraphics [width=5.5cm]{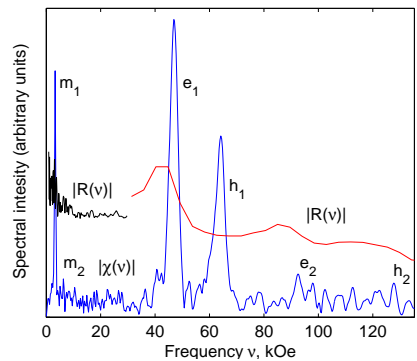}
\caption{Spectral intensity of dHvA oscillations of susceptibility
$\left| \protect\chi (\protect\nu )\right| $ and of SdH oscillations
of magnetoresistance $\left| R(\protect\nu )\right| $. Peaks
$m_{1,2}$, $e_{1,2} $, $h_{1,2}$ correspond to the 1st and 2nd
harmonics of oscillations from minority electrons, majority
electrons and majority holes. The low and high frequency plots of
$\left| R(\protect\nu )\right| $ are obtained from different sets of
experimental data.} \label{DiracFigFourrier}
\end{figure}

Fig.\ref{DiracFigExperiment} shows the measured magnetic susceptibility $%
\chi =dM/dH$ and the oscillating part of resistance $\Delta R$
(after substraction of the large polynomial background $R_{0}(H)$)
as a function of the inverse magnetic field $H^{-1}$ in the high-
and low-field regions. In agreement with previous experiments
\cite {Brandt1988,Soule1964,Williamson1965,Woollam1971}, $\chi
(H^{-1})$ is a superposition of at least three oscillating
contributions. This can
be seen in Fig. \ref{DiracFigFourrier} where both principal peaks $m_{1}$, $%
e_{1}$ and $h_{1}$ and their second harmonic counterparts $m_{2}$ $e_{2}$
and $h_{2}$ in spectral intensity of Fourier transformed susceptibility $%
\left| \chi (\nu )\right| $ are plotted. At the same time, only the $m$ and $%
e$ peaks are seen in the spectral intensity of resistance $\left|
R(\nu )\right| $. In other words, the SdH $h$-oscillations are
strongly damped. The corresponding oscillation frequencies $\nu_i$
and their assignment to the different groups of carries which we
justify below, are given in
Table \ref{kopelTabFactors}. Note that unlike the widely accepted result $%
\nu_h<\nu_e$ \cite{Woollam1971}, the hole frequency $\nu_h$ in our sample is
\textit{higher} then the electron frequency $\nu_e$.

\begin{table}[tbp]
\caption{ Frequencies $\protect\nu_i $ and phases $\protect\phi_i$, phase
factors $\protect\mu_i$, $\protect\gamma_i$, $\protect\delta_i$ and
assignment of quantum oscillations in graphite. }
\label{kopelTabFactors}\centering \vspace{5mm}
\begin{tabular}{lccrccl}
\hline\hline
& $\nu_i $ (kOe) & $\phi_i$ & $\mu _{i}$ & \ \ $\gamma _{i}$ \  & $\delta
_{i}$ \  & Assignment \\ \hline
minority m & $3.28$ & $0$ & $-1$ & $1/2$ & $0$ & \ normal h, 2D \\
majority e & $46.8$ & $0.75\pi$ & $1$ & $1/2$ & $-1/8$ & \ normal e, 3D \\
majority h & $64.1$ & $\pi$ & $-1$ & $0$ & $0$ & \ Dirac h, \ \ 2D \\
\hline\hline
\end{tabular}
\end{table}

\begin{figure}[!bt]
\centering
\includegraphics [width=6cm]{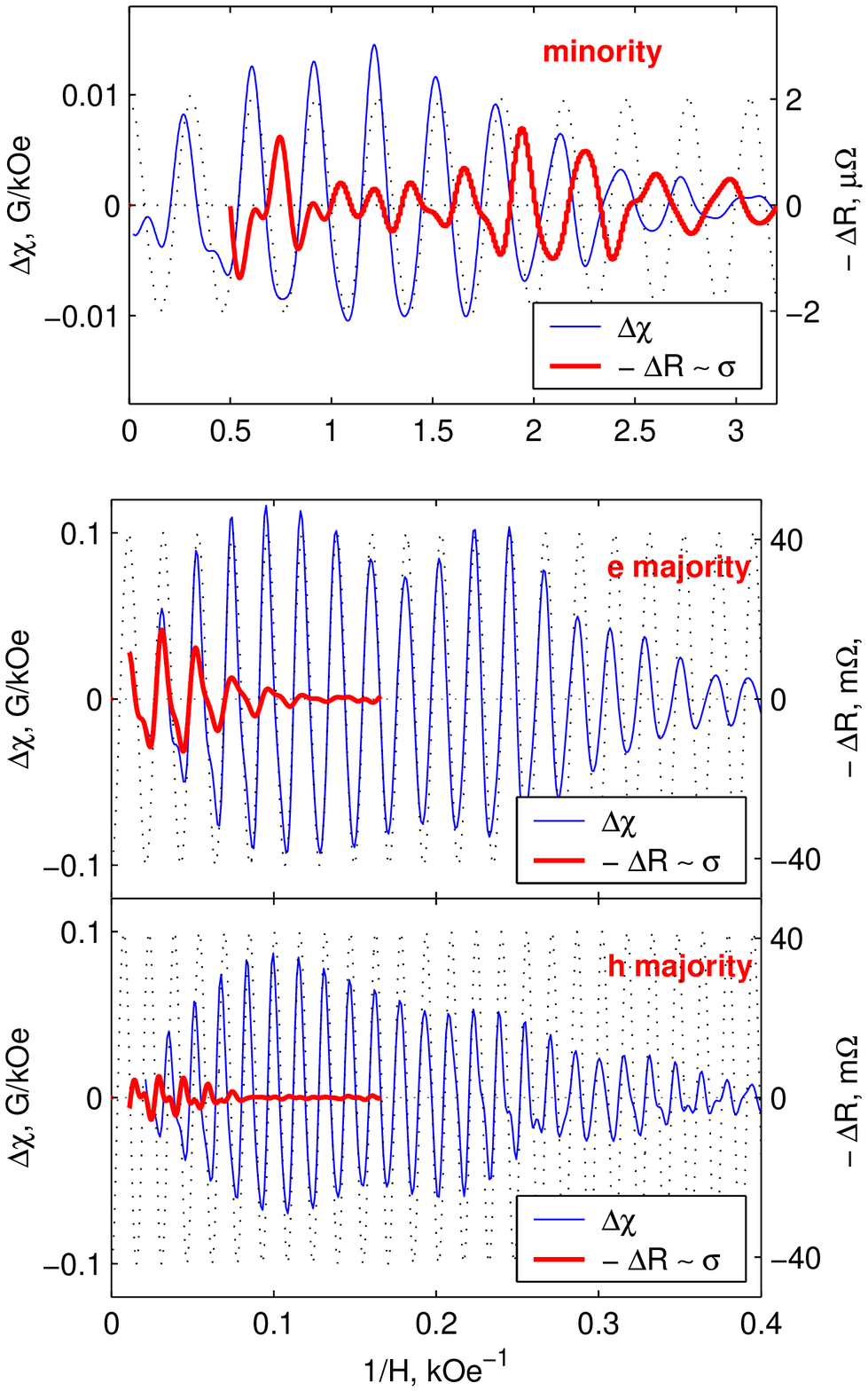}
\caption{Quantum oscillations of susceptibility $\Delta
\protect\chi (H^{-1}) $ and of resistance $\Delta R(H^{-1})$ for
different groups of carriers, obtained after two-harmonic
band-pass filtering of experimental data. Dot-lines show the
one-harmonic phase fit using the phase factors  specified in Table
\ref {kopelTabFactors}. The sign of $\Delta R$ is inverted to
recover the  oscillating part of the conductivity $\Delta
\protect\sigma$.} \label{DiracFigFiltre}
\end{figure}

We decomposed the measured signals $\chi (H^{-1})$ and $R(H^{-1})$
onto individual $m$-, $e$- and $h$-oscillations, applying the
frequency filtering with selective pass-bands about the 1st and
2nd harmonics of the corresponding resonant frequencies $\nu
_{i}$. As shown in Fig. \ref{DiracFigFiltre}; the results
demonstrate the generic behavior for quantum oscillations: an
initial growth of the high-field amplitude (low Landau levels)
followed by a low-field Dingle attenuation $\sim e^{-A_{i}/H}$.
 The low-intensity SdH h-oscillation are recovered from the noisy background of $%
R(\nu )$ around $\nu _{h}$. The sign of $\Delta R$ in Fig. \ref
{DiracFigFiltre} is reversed in order to recover the behavior of oscillating
part of the conductivity $\Delta \sigma =\Delta (\rho ^{-1})\approx -\Delta
\rho /\rho _{0}^{2}\sim -\Delta R$.

To proceed with the phase detection, we analyze how the nature of
the carriers influences the phase of the quantum oscillations, by
considering the  quasi 2D spectrum appropriate for graphite:
\begin{equation}
\varepsilon (\mathbf{p})=\varepsilon _{\perp }(p_{\perp })-2t\cos
p_{z}d \pm \mu_B H, \label{q2D}
\end{equation}
where the perpendicular dispersion $\varepsilon _{\perp }(p_{\perp
})$ can be either of the massive (parabolic) or of the Dirac
(linear) type (see Table \ref{kopelTabSpectra}); $\pm \mu_B H$ is
the Zeeman splitting ($\mu_B=e \hbar /2mc$).

The original theory of dHvA oscillations of Lifshitz and Kosevich
\cite
{Lifshitz1955} was developed for 3D metals with an arbitrary dispersion $%
\varepsilon (p)$, applied to the spectrum (\ref{q2D}) when the
energy spacing $\hbar \omega _{c}$ between Landau levels at the FS
is smaller then the characteristic dispersion $t$ along $z$. The
other limit of almost 2D electrons was studied quite recently and
the general expression that incorporates both $\hbar \omega
_{c}\ll t$ and $\hbar \omega _{c}\gg t$ limits was derived in
\cite{ChampelMineev2001} for the case of parabolic dispersion of
$\varepsilon _{\perp }(p_{\perp })$. It is straightforward to
generalize the calculations of \cite{ChampelMineev2001} for the
case of \textit{arbitrary } dispersion, using as in
\cite{Lifshitz1955} the electron orbit area $S(\varepsilon
,p_{z})=\pi p_{\perp }^{2}(\varepsilon ,p_{z})$ instead of
$\varepsilon _{\perp }(p_{\perp })$ and  the Bohr-Sommerfeld
semiclassical quantization:
\begin{equation}
S(\varepsilon ,p_{z})=(n+\gamma )2\pi \hbar \frac{eH}{c},  \label{Bohr}
\end{equation}
where the factor $0\leq \gamma <1$ is related to the topology of
the FS: $\gamma =1/2$ for parabolic dispersion and $\gamma =0$ in
the Dirac case \cite{Mikitik99}.

Reproducing the calculations of \cite{ChampelMineev2001} in terms of $%
S(\varepsilon ,p_{z})$ we get for the oscillating part of the
magnetization:
{\setlength\arraycolsep{1pt}
\begin{eqnarray}
\Delta M &=&-\frac{4}{(2\pi )^{2}}\frac{1}{\hbar d}\frac{e}{\hbar c}\frac{S}{%
dS/d\varepsilon }\sum_{l=1}^{\infty }\frac{1}{l}\frac{\lambda
l}{\sinh
\lambda l}e^{-\frac{\Gamma }{\hbar \omega _{c}}2\pi l}   \label{Formula}\\
\times J_{0}&(&2\pi l\frac{2t}{\hbar \omega _{c}})\sin \left( 2\pi
l\left[ \frac{c}{e\hbar }\frac{S(\varepsilon )}{2\pi H}-\gamma
\right] \right) \cos \left( 2\pi l \frac{\mu_B H}{\hbar \omega_c}
\right), \notag
\end{eqnarray}
where $\Gamma $ is the impurity width of the Landau level and parameters
\begin{equation*}
S(\varepsilon )=S(\varepsilon ,{\pi /2d}),\quad \lambda =\frac{\pi cT}{%
e\hbar H}\frac{dS}{d\varepsilon }\ll 1,\quad \omega _{c}=\frac{eH}{c}\frac{%
2\pi }{dS/d\varepsilon }
\end{equation*}
are given in Table \ref{kopelTabSpectra} for the normal and Dirac fermions.

\begin{table}[tbp]
\caption{Spectra, Landau quantization, areas of the
quasi-classical electronic orbits and parameters of the dHvA
oscillations for the massive (Normal) and masseless (Dirac)
fermions} \label{kopelTabSpectra}\centering \vspace{5mm}
\begin{tabular}{lll}
\hline\hline
& Normal & Dirac \\ \hline
$\varepsilon (\mathbf{p})$ & ${p_{\perp }^{2}}/{2m_{\perp }}$ & $\pm v |{%
p_{\perp }| }$ \\
$\varepsilon _{\perp }(n)$ \ \ \ \ \ \  & $\left({e\hbar }/{m_{\perp }c}%
\right)H(n+\frac{1}{2})$ \ \ \ \ \  & $\pm (2v^{2}e\hbar /c)^{1/2}(Hn)^{1/2}$
\\
$S(\varepsilon)$ & $2\pi m_{\bot }\varepsilon $ & $\pi \varepsilon ^{2}/v^{2}
$ \\
$\omega _{c}$ & $eH/cm_{\bot }$ & $ev^2H/c \varepsilon $ \\
$\lambda $ & $2\pi ^{2}cm_{\bot }T/e\hbar H$ \ \ \ \ \ \  & $%
2\pi^2cT\varepsilon/\hbar e v^2 H$ \\ \hline\hline
\end{tabular}
\end{table}

Calculated with respect to the band origin (at $p_{z}=0$) the
chemical potential $\epsilon $ equilibrates the oscillating
Fermi-levels of different groups of carriers and therefore acquires
the field dependence that was
shown \cite{Champel2001} to be important for the very clean 2D systems with $%
\omega _{c}\gg \Gamma $ and in the ultra-quantum limit when only low
Landau levels ($n\sim 1)$ are occupied. In the opposite case,\emph{\
} we neglect this dependence and assume that $\epsilon =\epsilon
_{F}$. Note, however, that beats in the majority oscillations (Fig.
\ref{DiracFigFiltre}) can be attributed to the conserving $\epsilon $
''cross-talk'' between e and h carriers \cite {Alexandrov1996}. We
neglect also the last oscillating spin factor in (\ref{Formula})
since the Zeeman splitting $\mu_B H$ in graphite is much smaller
than the distance between Landau levels ${\hbar \omega_c}$
\cite{Woollam1971}. This tiny splitting feature is observed only in
the high-field e-oscillations in Fig.\ref{DiracFigFiltre}.

Equation (\ref{Formula}) includes both the 3D Lifshitz-Kosevich
limit \cite
{Lifshitz1955} when $\zeta =2\pi l\frac{2t}{\hbar \omega _{c}}\gg 1$, $%
J_{0}(\zeta )\approx (2/\pi \zeta )^{1/2}\cos (\zeta -\pi /4)$ and
pure 2D limit when $\zeta \ll 1$, $J_{0}(\zeta )\approx 1$. In the
case of 2D Dirac fermions it reduces to the result obtained in
Ref. \cite{Sharapov2003} , whereas in the case of massive fermions
the result of \cite {ChampelMineev2001} is recovered. Analyzing
only the oscillating part of $\Delta M$, we find that the lst
harmonic of the magnetic susceptibility $\Delta \chi =d(\Delta
M)/dH$ oscillates as:
\begin{equation}
\Delta \chi _{l}\sim \mu \cos \left( 2\pi \left[ \frac{\nu }{H}l-\gamma
l+\delta \right] \right)  \label{Schem}
\end{equation}
where the factor $\mu =sign(\epsilon )$ is equal to $+1$ for the electrons
and to $-1$ for the holes. The topological index $\gamma $\ (see Eq. (%
\ref{Bohr})) is equal to $1/2$\ for massive fermions and is $0$\
for Dirac fermions. The factor $\delta $ reflects the curvature of
the FS in the z-direction and changes from $0$ for a quasi-2D
cylindrical FS when $\hbar \omega _{c}\ll t $ to $\pm \pi /8$ for
a corrugated 3D FS when $\hbar \omega _{c}\gg t$ ( $\pm $
corresponds to the contribution from minimal/maximal cross
section).

\begin{figure}[!bt]
\centering
\includegraphics [width=7cm]{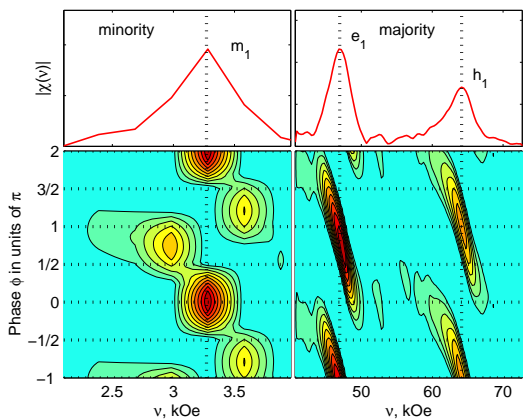}
\caption{Contour plot of the positive part of the phase-shift function $K(%
\protect\phi ,\protect\nu)=\textsf{Re} e^{i \protect\varphi }\protect\chi (%
\protect\nu )$ for minority and majority carries. Position of maxima of $K(%
\protect\phi ,\protect\nu)$ determines the oscillation frequencies $\protect%
\nu_i$ and phases $\protect\phi_i$ for different groups of carriers. Upper
panel presents the corresponding spectral intensity $| \protect\chi (\protect%
\nu )|$.}
\label{DiracFig2D}
\end{figure}

In order to determine factors $\mu_i$, $\gamma_i$ and $\delta_i$
for each group of presented in Fig. \ref{DiracFigFiltre}
oscillations we extract the phase of their 1st harmonics $\varphi
_{i}$, analyzing Fourier-transformed susceptibility $\chi (\nu )$
in the vicinity of oscillation frequencies $\nu _{i}$. Because the
phase information can not be extracted from the spectral intensity
plot $\left| \chi (\nu )\right| $ (Fig. \ref{DiracFigFourrier}),
we developed the method of the 2D phase-frequency analysis that is
free from the above mentioned phase/frequency uncertainty.
 For its illustration we assume  that close to the
resonance frequency $\nu_0$ the Fourier-transformed susceptibility
has the Gaussian-like profile $\chi (\nu) \sim
e^{i\varphi_0}e^{-\beta(\nu-\nu_0)^2}$. Constructing now \textit{the
phase-shift function} $K(\varphi,\nu)=\textmd{Re}e^{-i\varphi }\chi
(\nu )$ that in our model case is equal to $ e^{-\beta(\nu-\nu_0)^2
} \cos(\phi-\phi_0)$ we can detect both the  frequency $\nu_0$ and
the phase  $\varphi_0$ simultaneously as the position of the maximum
of $K(\varphi,\nu)$ in the  plane $\phi$-$\nu$.

The phase-shift function for the dHvA oscillations in graphite is
shown in Fig. \ref{DiracFig2D}. Determination of maxima of
$K(\varphi ,\nu )$ gives the collected in Table
\ref{kopelTabFactors} oscillation frequencies $\nu _{i}$ and their
phases $\varphi _{i}$ more precisely than the previous
determination: ($\phi_e=\phi_h={0.75}\pi$ in \cite {Soule1964} and
$\phi_h=0.76\pi$, $\phi_e=0.64\pi$ in \cite{Williamson1965}).

By analyzing the oscillations of $\Delta \sigma $\ and $\Delta
\chi$ (Fig. \ref{DiracFigFiltre}), and taking into account the
relation $\Delta \sigma \sim \mu |m|H^{2}\Delta \chi $\
\cite{Landau10}, we conclude that the in- and out-of-phase
behavior corresponds to electrons and holes with $\mu =+1$ and
$\mu =-1$, respectively. For h-carriers this analysis was
independently supported by a comparison between the longitudinal
resistance oscillations $\Delta R$ and of the Hall resistance
$\Delta R_{H}$ (not shown):  for holes, minima in $\Delta R$
should correspond to minima in $\Delta R_{H}$ \cite{Woollam1971},
and this is what we observed.

Knowledge of $\phi_i $ and $\mu_i $ allows the unambiguous
determination of the factors $\gamma_i $ ($\gamma_i =1/2$ or $0$)
and $\delta_i $ ($|\delta_i |<1/8$) that  are interrelated as:
\begin{equation}
\phi_i=\pi\left(\textrm{sign}\; \mu_i-2\gamma_i l + 2 \delta_i
\right) \label{Phasrel}
\end{equation}
The analysis of the 1th harmonic parameters $\mu_i$, $\gamma_i$
and $\delta_i$,  given in Table \ref{kopelTabFactors} for each
group of carriers, led us to the following conclusions.

(i) The minority carriers are holes with a 2D massive parabolic
spectrum.

(ii) The majority electrons have the parabolic spectrum with 3D
$p_{z}$ -dispersion.

(iii) The majority holes are 2D Dirac fermions.

The identification of Dirac fermions which can be responsible for
unusual strongly-correlated electronic phenomena in graphite
\cite{Kopelevich2003PRL,Kopelevich2003ASS} is the principal result
of this work.

The method proposed here for the
two-dimensional phase-frequency analysis allows the efficient
phase definition in any quantum oscillation phenomena, including
those in low-dimensional organic conductors, in the mixed state of
superconductors and in 2D quantum Hall semiconductors.

The work was supported by FAPESP and CNPq Brazilian scientific
agencies. I.L. thanks J.-L. Dellis for discussion of computation
problems.

\end{document}